# When Text Simplification Is Not Enough: Could a Graph-Based Visualization Facilitate Consumers' Comprehension of Dietary Supplement Information?

Xing He[1a], Rui Zhang[2,3a], Jordan Alpert[4], Sicheng Zhou[3], Terrence J Adam[2,3], Aantaki Raisa[4], Yifan Peng[5], Hansi Zhang[1], Yi Guo[1,6], Jiang Bian[1,6b]

[1] Health Outcomes & Biomedical Informatics, University of Florida, Gainesville, Florida, USA

[2] Department of Pharmaceutical Care & Health Systems, University of Minnesota, Minnesota, USA

[3] Institute for Health Informatics, University of Minnesota, Minnesota, USA

[4] Department of Advertising, College of Journalism and Communications, University of Florida, Gainesville, Florida, USA

[5] Department of Population Health Sciences, Weill Cornell Medicine, New York, USA

[6] Cancer Informatics Shared Resource, University of Florida Health Cancer Center, Gainesville, Florida, USA

**Email**: Xing He – hexing@ufl.edu; Rui Zhang – zhan1386@umn.edu; Jordan Alpert – jordan.alpert@ufl.edu; Sicheng Zhou – zhou1281@umn.edu; Terrence J Adam – adamx004@umn.edu; Aantaki Raisa – a.raisa@ufl.edu; Yifan Peng – yip4002@med.cornell.edu; Hansi Zhang – hansi.zhang@ufl.edu; Yi Guo – yiguo@ufl.edu; Jiang Bian – bianjiang@ufl.edu

[a] Xing He, MS and Rui Zhang, PhD contributed equally, co-first authors

[b] **Corresponding author**: Jiang Bian, PhD; bianjiang@ufl.edu.

Affiliation: Health Outcomes & Biomedical Informatics, University of Florida

Address: 2197 Mowry Road, 122 PO Box 100177 Gainesville, FL 32610-0177

Phone Number: (813)573-3122

**Word Count:** ~4000

# Abstract

**Background:** Dietary supplements are widely used. However, dietary supplements are not always safe. For example, an estimated 23,000 emergency room visits every year in the United States were attributed to adverse events related to dietary supplement use. With the rapid development of the Internet, consumers usually seek health information including dietary supplement information online. To help consumers access quality online dietary supplement information, we have identified trustworthy dietary supplement information sources and built an evidence-based knowledge base of dietary supplement information—the integrated DIetary Supplement Knowledge base (iDISK) that integrates and standardizes dietary supplement related information across these different sources. However, as information in iDISK was collected from scientific sources, the complex medical jargon is a barrier for consumers' comprehension.

**Objective:** To assess how different approaches to simplify and represent dietary supplement information from iDISK will affect lay consumers' comprehension.

**Methods:** Using a crowdsourcing platform, we recruited participants to read dietary supplement information in four different representations from iDISK: (1) original text, (2) syntactic and lexical text simplification, (3) manual text simplification, and (4) a graph-based visualization. We then assessed how the different simplification and representation strategies affected consumers' comprehension of dietary supplement information in terms of accuracy and response time to a set of comprehension questions.

**Results:** With responses from 690 qualified participants, our experiments confirmed that the manual approach, as expected, had the best performance for both accuracy and response time to the comprehension questions, while the graph-based approach ranked the second outperforming other representations. In some cases, the graph-based representation outperformed the manual approach in terms of response time.

**Conclusions:** A hybrid approach that combines text and graph-based representations might be needed to accommodate consumers' different information needs and information seeking behavior.

## Lay Summary

Dietary supplements are widely used but not always safe. Consumers often seek health information including dietary supplement information online. The integrated DIetary Supplement Knowledge base (iDISK) was created, integrating trustworthy dietary supplement information across scientific sources, to help consumers access quality online dietary supplement information. However, the complex medical jargon from scientific sources is a barrier to consumers' comprehension, where text simplification combined with other alternative information presentation methods can potentially be useful. We developed three different information representation approaches: (1) syntactic and lexical text simplification, (2) manual text simplification, and (3) a graph-based visualization, and assessed how these different approaches to simplify and represent dietary supplement information from iDISK will affect consumers' comprehension. To do so, we used a crowdsourcing platform to recruit participants to read dietary supplement information in different representations, and then assessed their ability to answer a set of comprehension questions in terms of accuracy and response time. Our experiment results show that the manual text simplification approach had the best performance for both accuracy and response time, while the graph-based approach ranked the second. A hybrid approach that combines text and graph-based representations might be needed to accommodate consumers' different information needs and information-seeking behavior.

## Introduction

More than 77% of Americans take dietary supplements based on the latest 2019 survey commissioned by the Council for Responsible Nutrition [1]. Vitamins and minerals continue to be the most popular dietary supplement products, where more than 76% of US adults have taken at least one dietary supplement in the past year [1]. The consumption of dietary supplement is generally high regardless of age groups: 70% of adults between the ages 18 – 34, 81% of adults between the ages 35 – 54, and 79% of adults ages over 55 [1]. Many dietary supplement consumers expressed overall confidence in the safety, quality, and effectiveness of dietary supplement use; however, dietary supplement are not always safe. For example, more than 15 million US adults are at risk for drug-supplement interactions (DSIs) or high-dose vitamins [2]. An estimated 23,000 emergency room visits every year in the US were attributed to adverse events related to dietary supplement use [3]. There is also increasing evidence that dietary supplement can interact with a wide range of prescription medications, resulting in adverse events [4]. Despite these safety concerns, there are significant information gaps on appropriate dietary supplement use for consumers. Further, dietary supplement consumption is not disclosed by patients to their physicians in 42.3% of all cases [5] and even lower rates of communication were noted with pharmacists [6].

The use of dietary supplement is often self-medicated, leading patients to seek relevant dietary supplement use information on their own [7]. The Internet is the first place for consumers to find health information [8,9]. Based on our recent study analyzing questions related to dietary supplement on Yahoo! Answer, we found consumers frequently seek information on dietary supplement usage, adverse effects, and addiction [10]. Currently, many online sites contain basic dietary supplement information, their therapeutic use, safety warnings, effectiveness, and information on dietary supplement related research studies. However, much of this online information consists of opinions, salesmanship, testimonials, and claims that are not evidence-based. Access to quality online health information has long been a concern [11]. In our prior study, we have identified trustworthy dietary supplement information sources such as product labels from the Dietary Supplement Label Database (DSLD) and patient educational information from Memorial Sloan Kettering Cancer Center (MSKCC) and built a evidence-based knowledge base of dietary supplement information—the integrated DIetary Supplement Knowledge base (iDISK) that integrates and standardizes dietary supplement related information across these different sources [12,13].

Nevertheless, as information in iDISK was collected from scientific sources (e.g., scientific literature and monographs written by clinicians and scientists), complex medical jargon is a barrier to consumers' comprehension of dietary supplement related health information, resulting in confusion and potentially inappropriate use of dietary supplement products. As only 12% of US adults are considered to have proficient health literacy [14], a number of government agencies and national programs (e.g., the Clear Communication Index from the Centers for Disease Control and Prevention [15] and the Clear Communication initiative at the National Institutes of Health [16]) recommended that health information content should be (1) written in plain language that is understandable to lay consumers, and (2) clear and simple, especially when developing content for people with limited literacy skills. Built on iDISK, we previously developed ALOHA—an interactive graph-based visualization platform to facilitate consumers' browsing and understanding of dietary supplement information in the iDISK [17], following a user-centered design process. The usability testing of ALOHA was acceptable (i.e., a System Usability Scale score of 64.4 ± 7.2) and most participants in the usability testing sessions thought that the graph-based visualization in ALOHA is a creative and visually appealing format to obtain health information. Nevertheless, it is not clear yet whether a graph-based visualization or simply text simplification (TS) will improve lay consumers' comprehension of evidence-based dietary supplement information in iDISK.

In the past, Amazon Mechanical Turk (MTurk), a web-based microtask crowdsourcing platform, where individuals perform human intelligence tasks online in exchange for payment, has been used to evaluate laypeople's comprehension of medical information. For example, Yu et al. (2013) employed MTurk to evaluate laypeople's comprehension of medical pictograms [18]. Lalor et al. (2018) utilized MTurk to assess patients' electronic health record note comprehension [19]. Cho et al. (2020) used MTurk to assess patient comprehension of radiology reporting templates and radiology colloquialisms [20].

In this study, we aimed to assess how different approaches to simplify and represent dietary supplement information from iDISK will affect lay consumers' comprehension. By using MTurk, we tested four different representations: (1) original scientific language,

(2) TS through manual curation, (3) TS through a hybrid simplification approach (i.e., syntactic simplification model followed by lexical simplification replacing medical jargons with terms from consumer health vocabulary [CHV]) [21], and (4) graph-based visualization, and assessed how the different simplification strategies affected consumers' comprehension of dietary supplement information in terms of accuracy and response time to the comprehension questions.

## Methods

This study was reviewed and approved as Exempt by the University of Florida Institutional Review Board under protocol number IRB202100602.

### Data Sources

We randomly selected 10 dietary supplement ingredients and extracted their information from the iDISK, which encompasses both a terminology of dietary supplement ingredients and a structured knowledge base of dietary supplement related information [13]. iDISK was developed through integrating essential dietary supplement information from four commonly used and trusted dietary supplement resources: the Natural Medicines Comprehensive Database (NMCD)—a commercial dietary supplement ingredient-level database, the "*About Herbs*" page on the MSKCC website, the DSLD—a database of dietary supplement product labels with over 76,000 dietary supplement products marketed in the US, and the Natural Health Products Database (NHP)—a database that covers natural health products with a product license issued by Health Canada. We initially extracted 3 sections of textual information about a dietary supplement ingredient from iDISK: (1) background (i.e., a summary of information about the ingredient such as its origination, uses, and constituent, extracted from NMCD, MSKCC, and NHP), etc., (2) mechanism of action (i.e., the mechanism by which an active substance produces an effect on a living organism or in a biochemical system, based on MSKCC), and (3) safety (i.e., a summary of the safety concerns such as adverse reactions associated with using the ingredient). Through discussions with the clinician on the team, we reached a consensus that the current information on the mechanism of action for a dietary supplement ingredient in iDISK is too granular and intended for health care professional use. Thus, we focused on the background and safety information in our consumer comprehension experiments. The length of the text for each dietary supplement ingredient varies, however, is typically too long (i.e., a couple of paragraphs) for a crowdsourcing experiment. Thus, we extracted a random paragraph (i.e., 3 to 5 continuous sentences) from each section of the dietary supplement information.

**Overall Study Design**

*Figure 1* shows the flow of our study design. We tested 4 different representations of the dietary supplement information: the original text, two TS strategies (i.e., manual curation and a hybrid syntactic with lexical TS strategy), and a graph-based strategy. *Table 1* shows an example of the 3 text-based representations, and *Figure 2* shows an example of the graph-based representation for the dietary supplement ingredient "Omega 3". We manually created one comprehension question for each of the two sections (i.e., background and safety) for each dietary supplement ingredient, presented the different representations of the dietary supplement information to participants recruited through MTurk, and assessed their comprehension of the dietary supplement information through their answers to the questions and response time. We also collected basic demographics of each participant (e.g., age groups, gender, race) as well as their health literacy level using the validated Newest Vital Sign (NVS) health literacy assessment tool developed by Pfizer [22].

*Table 1.* An example of the 3 text-based representations for the background section of "Omega 3."

| Original | Manual | Syntactic + Lexical (Selected Content)[a] | Question |
|---|---|---|---|
| A type of polyunsaturated fatty acid (PUFA) **derived mainly** from fish oil, omega-3 fatty acids are used as a **dietary** supplement for depression, to lower cholesterol, and to reduce the risk of heart attack. **Data from a randomized trial** suggest that omega-3 may be useful in **reducing the risk of progression** to psychiatric disorders and as a safe preventive | A type of polyunsaturated fatty acid (PUFA) **developed mostly** from fish oil, omega-3 fatty acids are used as a **food** supplement for depression, to lower cholesterol, and to cut the risk of heart attack. **Research suggests** that omega-3 may be useful in **cutting the risk of growth** to psychiatric problems and as a safe preventive measure in young adults at risk for psychotic health | A types of polyunsaturated fatty acids are used by as a dietary supplement for **mental depression**, **togo** lower cholesterol

Data from a randomized **clinical trials** suggest that omega 3 fatty acid **may be usage in reduced** the risk of progression to **mental illness** at risk for **mental disorder** | Do omega-3 fatty acids increase cholesterol and increase the risk of heart attack?

Answer: No |

| | | | |
|---|---|---|---|
| measure in young adults at risk for psychotic conditions. Omega-3 fatty acid supplementation lowers cholesterol and **may reduce recurrence in patients with a history of stroke.** | problems. Omega-3 fatty acids lowers cholesterol and may **cut the risk of** stroke for **patients who had a stroke in the past.** | omega 3 fatty acid dietary supplementation may reduce **recurring** in patient with a **medical history** of stroke | |

[a] A machine learning-based synaptic text simplification model was run first, and medical jargons were then replaced based the Consumer Health Vocabulary resource. Due to page limit, only a selected set of results is shown here.

*Manual Text Simplification and Question Generation*

Two co-authors (JA and AR) with background in health communication manually simplified the original text with the help of a commercial product—Health Literacy Advisor (HLA) [23]. HLA is an interactive health literacy tool that that highlights complex health terms such as words with more than three syllables. Plain language replacements are suggested based on various validated readability indexes, such as Fry, ARI, Precise SMOG Index, FORCAST Readability Grade, and Flesch Reading Ease Score. HLA has been used to improve the readability of health documents and health education materials [24,25], improve clinical summaries for patients [26], refine messages for health interventions [27,28], and for the evaluation of educational programs for cancer survivors [29]. However, HLA suggestions did not account for the impact on sentence structure and overall readability. For instance, sentences incorporated with HLA suggestions were sometimes grammatically incorrect or the meaning of the sentence was changed. Thus, the two co-authors manually edited the HLA-simplified text making grammatical edits and replacing HLA suggestions if the original meaning was lost. The authors also searched the Internet and used verified government health websites to find common terms for medical and scientific jargon not replaced by HLA. The most extensive edits involved re-arranging sentences within each description to produce a narrative style summary (i.e. changing the order of subject-verb-predicate or making passive sentences active). Based on elements that were changed during the manual edits, for each ingredient, we also generated two questions with gold-standard yes or no answers—one for each of the background and safety sections, respectively. A pharmacist (TA) reviewed the manual descriptions as well as the questions to ensure that the original meaning of the text was retained.

### Syntactic and Lexical Text Simplification Approach

We applied both syntactic simplification and lexical simplification on the original texts. For syntactic simplification, we used the iSimp [30] tool (developed by YP) to process the original texts at the document level. The iSimp is a sentence simplification system designed to detect various types of clauses and constructs used in a complex sentence and produce multiple simple sentences while maintaining both coherence and meaning of the communicated message. It first tokenizes the text into a sequence of non-overlapping chunks and uses recursive transition networks to detect simplification constructs. And then, it generates simplified sentences by combining various simplification constructs.

Currently, iSimp can detect six major types of simplification constructs, including coordination, relative clause, apposition, introductory phrase, subordinate clause, and parenthetical element.

After syntactic simplification, we did lexical simplification by replacing medical jargons with mapped lay consumer terms from the Consumer Health Vocabulary (CHV) [31] in the Unified Medical Language System (UMLS) [32]. CHV is a collection of terms found to best represent the medical concepts for consumers and mapped to corresponding professional terms. Previous studies have shown that CHV terms were more comprehensible by patients when compared with their professional synonyms [21]. The lexical simplification includes three main steps: (1) detecting the potential medical jargons and locating their positions in the sentences, (2) identifying the UMLS Concept Unique Identifiers (CUI) of potential matched CHV terms, and 3) replacing the medical jargons with the CHV terms identified by CUI. The first two steps were completed using MetaMap [33], which analyzes the words in the text and matches the candidate words to UMLS vocabularies.

### *Graph-based Visualization of Dietary Supplement Information*

Based on our prior work (i.e., ALOHA—an interactive graph-based visualization platform to facilitate browsing of dietary supplement information in the iDISK) [17], we further developed a graph-based visualization to represent the dietary supplement information. The visualization relied on an open-source web-based graph visualization framework – InteractiveGraph [34]. To generate the visualization, we manually extracted semantic triples (e.g., "omega-3 fatty acids" – "are used as" – "food supplements" in [subject]-<predicate>-[object]) from the original text. Then we built the visualization by transforming the subjects/objects to nodes and the predicates (or relations) to links between the subject and object nodes. *Figure 2* shows an example of the graph-based representation for the background section of "Omega 3."

*Crowdsourcing Experiments to Assess Consumers' Comprehension of Dietary Supplement Information Using Different Representations*

To assess how these different simplification strategies will affect consumer's comprehension of dietary supplement information, we designed a web-based tool, namely Simplified Text Understanding Test (STUT), to facilitate the MTurk experiment. STUT consists of 5 parts: (1) a brief onboarding video tutorial about how to use STUT, (2) four demographic questions (i.e., age group, gender, ethnicity, and race), (3) the NVS health literacy assessment, (4) the simplified representation and corresponding questions, and (5) the reward code page. The NVS (*Figure 3*) health literacy assessment tool is a brief health literacy screening tool, where participants read a food nutrition label and answer six questions. Each of the correct answers earns one point, and the final NVS score, ranging from 0 to 6, is categorized into limited (0–1), marginal (2–3), or adequate (4–6) health literacy. The NVS was originally developed as an interviewer-administered health literacy assessment tool. Prior studies indicated that a computerized form of NVS assessment performed as well as the interviewer-administered version for assessing health literacy levels [35].

In MTurk, participants needed to read a simple description of the task before launching STUT to work on a task, where the onboarding tour (i.e., a 30-second video) would popup automatically first. The participants were required to complete the onboarding tour, the demographic questions, and the NVS health literacy assessment questions. As shown in *Figure 4*, one of the different representations (i.e., simplified text or graph-based visualization) appears on the left side of the page, and the corresponding comprehension question appears on the right side. After completing all questions (i.e., 10 questions per participant), the participant received a reward code to claim the incentives on MTurk. We designed the STUT tool so that it logged each participant's interactions with the tool and captured the amount of time they spent on each section.

Before running the formal experiments, we conducted a pilot study to estimate the adequate incentives for MTurk workers and the necessary sample size needed to detect the effect of how different representations would impact consumers' comprehension of dietary supplement information. We first released a set of assignments with the original text only, where we incrementally increased the incentives from 25 cents to 1 dollar and assessed how long it took for each assignment to be completed on MTurk. We then released 20 assignments (i.e., allowing 20 participants to complete the same task) on MTurk for both original text and manually simplified text, respectively, with an incentive of $1 per assignment. In this test run, we collected 20 valid responses from the original text group, and 19 valid responses from the manual simplified text group. The accuracies of their responses to the comprehension questions were 85.5% and 97.3%, for the original and manual text groups, respectively; while the mean time spent on completing the tasks

(i.e., 10 questions per task) were 387.5 seconds and 283.2 seconds, respectively. Based on the accuracy measures, we estimated the appropriate sample size for detecting a difference between two proportions (i.e., 85.5% vs. 97.3%) at 95% confidence level with 80% power is 85.

Thus, in the subsequent formal experiments, we released 100 assignments for each question group (i.e., 5 ingredients and 10 questions, where each ingredient would have two questions—one related to the "background" section and one related to the "safety" section) for each representation, and each participant was paid $1 for successfully completing a task (i.e., one question group, 10 questions). We also included a number of validation tests to avoid participants using automated scripts (i.e., bots) to complete the tasks or to detect low quality responses due to participants not paying attention to the tasks. Data from participants who did not pass all the validation tests were excluded; and participants would not receive incentives for disqualified responses. Two multiple-choice validation questions (e.g., "Tom is a grade 2 student. He is good at math. Tom likes painting, swimming and eating apple," with a question "Does Tom like eating apple?") were added and mixed with the other task questions in each assignment as validation tests. As an additional quality control mechanism, we also discarded responses that were completed in less than 90 seconds or larger than 1,000 seconds. We also implemented mechanisms to avoid multiple submissions from the same participant (i.e., thus, in our final dataset, each participant is only allowed to answer one question group, as being more familiar with the tasks may temper the integrity of the comprehension

tests). Finally, we set a required qualification in MTurk, so that the participants had to be located within the United States.

## Results

*Table 2* shows the basic demographics and the health literacy levels of the participants with qualified responses. The majority of the participants were less than 45 years old; and the gender distribution of the participants was fairly even across the four representation groups. A one-way ANOVA analysis was conducted for the average health literacy scores across the four groups of participants for the different representations. The test showed no significant difference ($P = .31$) among the four groups of participants' average health literacy scores.

*Table 2.* The basic demographic information of the participants for each representation.

|  | Original | Manual | Syntactic + Lexical | Graph |
|---|---|---|---|---|
| # of valid responses | N=180 | N=174 | N=165 | N=171 |
| Age | | | | |
| < 45 | 129 (71.7%) | 129 (74.1%) | 127 (77.0%) | 126 (73.7%) |
| 45 - 64 | 44 (24.4%) | 39 (22.4%) | 31 (18.8%) | 42 (24.6%) |
| ≥ 65 | 7 (3.9%) | 6 (3.4%) | 7 (4.2%) | 3 (1.8%) |
| Gender | | | | |
| Male | 93 (51.7%) | 75 (43.1%) | 89 (53.9%) | 81 (47.4%) |
| Female | 86 (47.8%) | 98 (56.3%) | 72 (43.6%) | 88 (51.5%) |
| Other | 1 (0.6%) | 1 (0.6%) | 4 (2.4%) | 2 (1.2%) |
| Health Literacy (HL) | | | | |
| NVS Score | 4.96±1.52 | 4.68±1.61 | 4.80±1.64 | 4.93±1.44 |
| Limited HL (0-1) | 12 (6.7%) | 14 (8.0%) | 14 (8.5%) | 7 (4.1%) |
| Marginal HL (2-3) | 16 (8.9%) | 22 (12.6%) | 18 (10.9%) | 21 (12.3%) |
| Adequate HL (4-6) | 152 (84.4%) | 138 (79.3%) | 133 (80.6%) | 143 (83.6%) |

*Table 3* shows the average rates of correct answers and the average time spent on answering each of the comprehension questions for each representation over the 10 dietary supplement ingredients (i.e., 20 comprehension testing questions total for each representation). *Table 4* shows the z-test of the average correct rates and *Table 5* shows the t-test of the average time spent across different representations. The manually simplified text representation performed consistently better than other approaches in terms of both the average rates of correct answers and the average time spent; while, the syntactic + lexical approach performed consistently the worst among the four representations. The graph-based representation performed better than the original text in all cases, although it performed slightly worse (85.7% vs. 92.7%) in terms of accuracy. The difference in the amount of time spent on each question, comparing the graph-based representation with the manually simplified text was not statistically significant (26.986 seconds vs. 25.432 seconds, $P = .30$).

*Table 3.* The average rates of correct answers and the average time spent across the four representations.

| | Average Accuracy (Rates of Correct Answers (%)) | Average Time Spent on Each Question (Seconds) |
|---|---|---|
| Original | 82.7±18.0 | 27.658±14.294 |
| Manual | 92.7±11.9 | 25.432±14.545 |
| Syntactic + Lexical | 70.9±23.2 | 35.762±17.437 |
| Graph | 85.7±16.2 | 26.986±13.413 |

*Table 4.* Z-test for the comparisons of average rates of correct answers across the four representations.

| | Comparisons of Average Rates of Correct Answers (%) | Z-test (P-value) |
|---|---|---|
| Original vs. Manual | 82.7±18.0 < 92.7±11.9 | -9.0557 ($P < .001$) |
| Original vs. Syntactic + Lexical | 82.7±18.0 > 70.9±23.2 | 8.2029 ($P < .001$) |
| Original vs. Graph | 82.7±18.0 < 85.7±16.2 | -2.4361 ($P = .01$) |
| Manual vs. Graph | 92.7±11.9 > 85.7±16.2 | 6.655 ($P < .001$) |

*Table 5.* T-test for the comparisons of average time spent across the four representations.

| | Comparisons of Average Time Spent on Each Question (Seconds) | T-test (P-value) |
|---|---|---|
| Original vs. Manual | 27.658±14.294 > 25.432±14.545 | 1.4518 ($P = .15$) |
| Original vs. Syntactic + Lexical | 27.658±14.294 < 35.762±17.437 | -4.6962 ($P < .001$) |
| Original vs. Graph | 27.658±14.294 > 26.986±13.413 | 0.4544 ($P = .65$) |
| Manual vs. Graph | 25.432±14.545 < 26.986±13.413 | -1.0455 ($P = .30$) |

*Figure 5* shows both the average rates of correct answers and the average amount of time spent across the four representations on each question. We also conducted z-test between the average rates of correct answers and t-test between the average time spent of different representations for each question. These results confirmed that the manually simplified text representation performed better than other representations in most of the questions, while the syntactic and lexical approach performed consistently the worst. The graph-based representation performed better than the original text representation in most questions with very few exceptions. When considering the average time spent, the graph-based representation worked consistently better than the original text representation, worse than the manually simplified version in 2 questions (i.e., questions 6—safety information about aloe gel and 13—background information about lemongrass), but better in other 2 cases (i.e., questions 8—safety information about acai and 15—background information about ginko biloba), and no statistically significant differences in all other questions.

## Discussion

The goal of this study was to assess how different simplification strategies of dietary supplement information from an evidence-based dietary supplement knowledge base, iDISK, will affect lay consumers' comprehension of the information. Specifically, we tested four different representations—one with the original text from iDISK, two text simplification strategies (i.e., manual and automated approaches), and one graph-based visualization. We assessed consumer's comprehension from two perspectives: accuracy

(i.e., correctly answered the comprehension question) and efficiency (i.e., time spent on reading the dietary supplement information and then answering the comprehension questions). From the accuracy perspective, our experiments indicated, as expected, that the manual text simplification approach had the best performance overall. Surprisingly, the automated syntactic and lexical hybrid text simplification approach performed significantly worst among the four representations, even when compared with the original text written in scientific language. The graph-based visualization approach performed, although slight worse than the manually simplified text, consistently better than the original text and syntactic + lexical simplified text representations. From the efficiency perspective, the manual text simplification and graph-based visualization approaches demonstrated similar performance, but significantly better than the original text and syntactic + lexical simplified text representations.

Syntactic and lexical-based text simplification approach had the worst performance in both efficiency and accuracy, possibly due to three reasons. First, the syntactic simplification divided a complex sentence into simpler, but a number of smaller sentences (e.g., from a complex sentence, "Patients take ginseng to improve athletic performance, strength and stamina, and as an immunostimulant." to a number of smaller sentences: "patient take ginseng to improve physical strength", "patient take ginseng to improve stamina", "patient take ginseng to improve as an immunostimulant", and "patient take ginseng to improve sport performance"). Even though the structure of the decomposed sentences was simpler, it will generate repeated phrases and words, which makes it take longer for the participants to read and find relevant key information.

Second, the syntactic simplification did not always produce grammatically correct sentences, confusing the participants. Third, compared to the manual simplification approach, the lexical simplification approach that replaced medical jargons using CHV terms did not always produce desired results. It was because of two likely reasons: (1) the CHV has not been updated since 2012 and does not have a good coverage of the consumer vocabulary for dietary supplement information, and (2) the CHV substitutions have replaced some of the keywords that also appeared in the comprehension questions, making it difficult for the participants to answer the comprehension questions.

The graph-based visualization ranked the second in the four representations, performed slightly worse in terms of accuracy, but in par in terms of efficiency compared with the manually simplified text representation. The reason could be multifold. First, the transformation of text from sentence to semantic triples only maintained key information words but could also have removed some of the contextual information. To some users, the contextual information might help them understand the information. Second, a close inspection of the 4 questions where the graph-based visualization performed better in 2, but worse in the other 2 questions, than the manual simplification in term of efficiency, revealed that the graph-based visualization can potentially help end users understand new or unfamiliar concepts (i.e., acai and ginkgo biloba) better than simple or common concepts (i.e., aloe gel and lemongrass). Further exploration is warranted to understand in what scenarios can graph-based visualization help lay consumers' comprehension of health information. Nevertheless, our current work did allude that a hybrid—leveraging both text simplification and graph-based visualization— approach can potentially be a better strategy to meet different consumers' different information needs and ultimately to improve consumers' comprehension of the information.

*Limitations*

A few limitations exist in our work. First, we only used one question to test participants comprehension of a paragraph, where correctly answering just one question may not reflect a good understanding of the paragraph reliably. A more reliable instrument is needed to assess users' comprehension. Second, the graph-based visualization we implemented was a simplified version of ALOHA [17], where we eliminated a number features (e.g., filtering by node type) to make it workable in a crowdsourcing setting. A more comprehensive testing of ALOHA with these convenient functions is needed in a future work. Last, the population recruited from MTurk is different (e.g., younger and more females) from the general population in the US. Careful considerations are needed before generalizing findings from crowdsourcing workers to the boarder population. For example, most participants in our study had adequate health literary. A more tailored approach is needed to customize a text-graph hybrid interface for low literacy individuals. The use of MTurk also has a number of other limitations (e.g., the presentence of bots and the potential of having "low quality" users that did not perform the tasks diligently). Nevertheless, a crowdsourcing platform such as MTurk gives researchers an easy access to a large number participants to produce preliminary data that can be used to generate more accurate hypotheses and design more rigorous experiments such as a randomized trials to answer these hypotheses.

*Conclusions*

Biomedical knowledge bases such as iDISK are booming with evidence-based information integrated from diverse scientific sources. Nevertheless, consumers of these scientific knowledge bases, especially patients but also clinicians, need supporting tools to help them identify, digest, and understand information relevant to their information needs. We tested 4 different information presentation strategies and found that a hybrid approach that combines text and graph-based representations might be needed to accommodate consumers' different information needs and information seeking behavior.

**Abbreviations**

DSI: drug-supplement interaction

DSLD: Dietary Supplement Label Database

MSKCC: Memorial Sloan Kettering Cancer Center

iDISK: integrated DIetary Supplement Knowledge base

TS: text simplification

MTurk: Mechanical Turk

CHV: Consumer Health Vocabulary

NMCD: Natural Medicines Comprehensive Database

NHP: Natural Health Products Database

NVS: Newest Vital Sign

HLA: Health Literacy Advisor

UMLS: Unified Medical Language System

CUI: Concept Unique Identifier

STUT: Simplified Text Understanding Test

NCCIH: National Center for Complementary & Integrative Health

ODS: Office of Dietary Supplement

NLM: National Library of Medicine

NIH: National Institutes of Health

**Funding Statement**

This work is supported by the National Center for Complementary & Integrative Health (NCCIH) and the Office of Dietary Supplements (ODS) grant number R01AT009457 (Zhang), the intramural program funds from the National Library of Medicine (NLM)/National Institutes of Health (NIH), NLM of NIH K99LM013001, funding from the National Cancer Institute (NCI) under award number R01CA246418, and the Cancer Informatics Shared Resource at the University of Florida Health Cancer Center. The content is solely the responsibility of the authors and does not represent the official views of the NIH.

**Contributorship Statement**

RZ, JA, YP, YG, and JB conceived and planned the experiments. XH, SZ and HZ carried out the experiments. RZ, JB, TA, AR, YG, JA and XH contributed to the interpretation of the results. XH, RZ, JB and SZ wrote the initial draft of the manuscript. All authors provided critical feedback and helped shape the research, analysis and manuscript.

## Competing Interests Statement

None declared.

## Data Availability

Data reported in the manuscript are available from the Dryad Digital Repository:

https://doi.org/10.5061/dryad.v9s4mw6v8

Integrated Dietary Supplement Knowledge Base (iDISK) is available from:

https://doi.org/10.13020/d6bm3v

**Figure Legends**

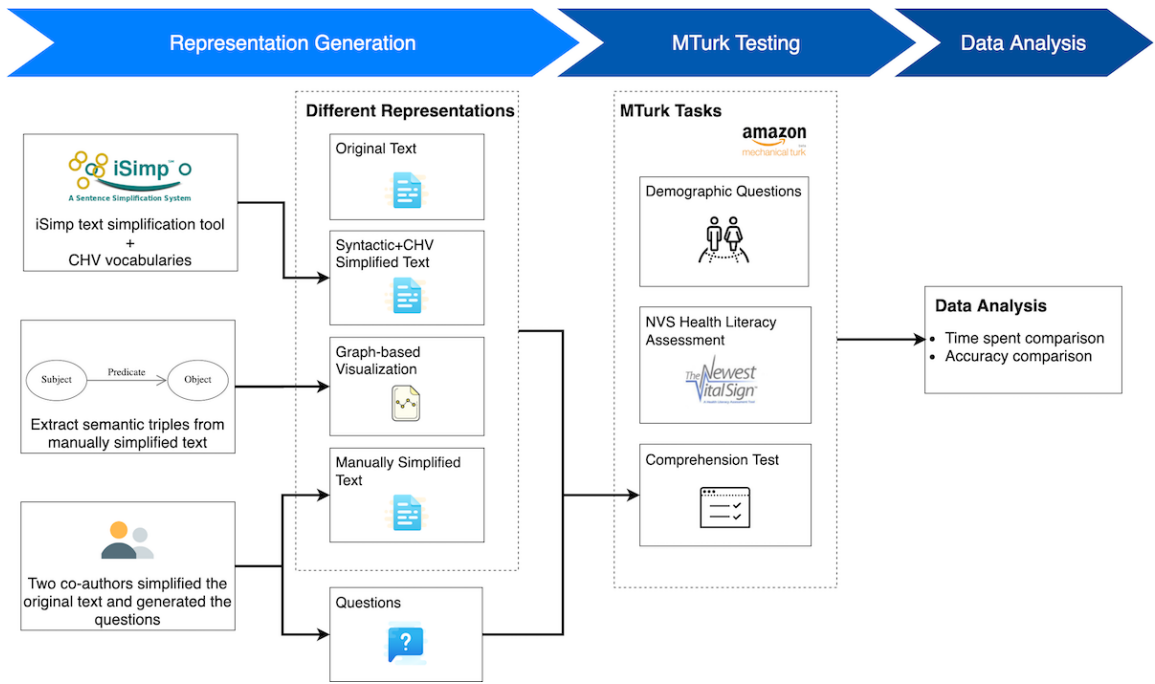

**Figure 1.** The overall design and flow of the study.

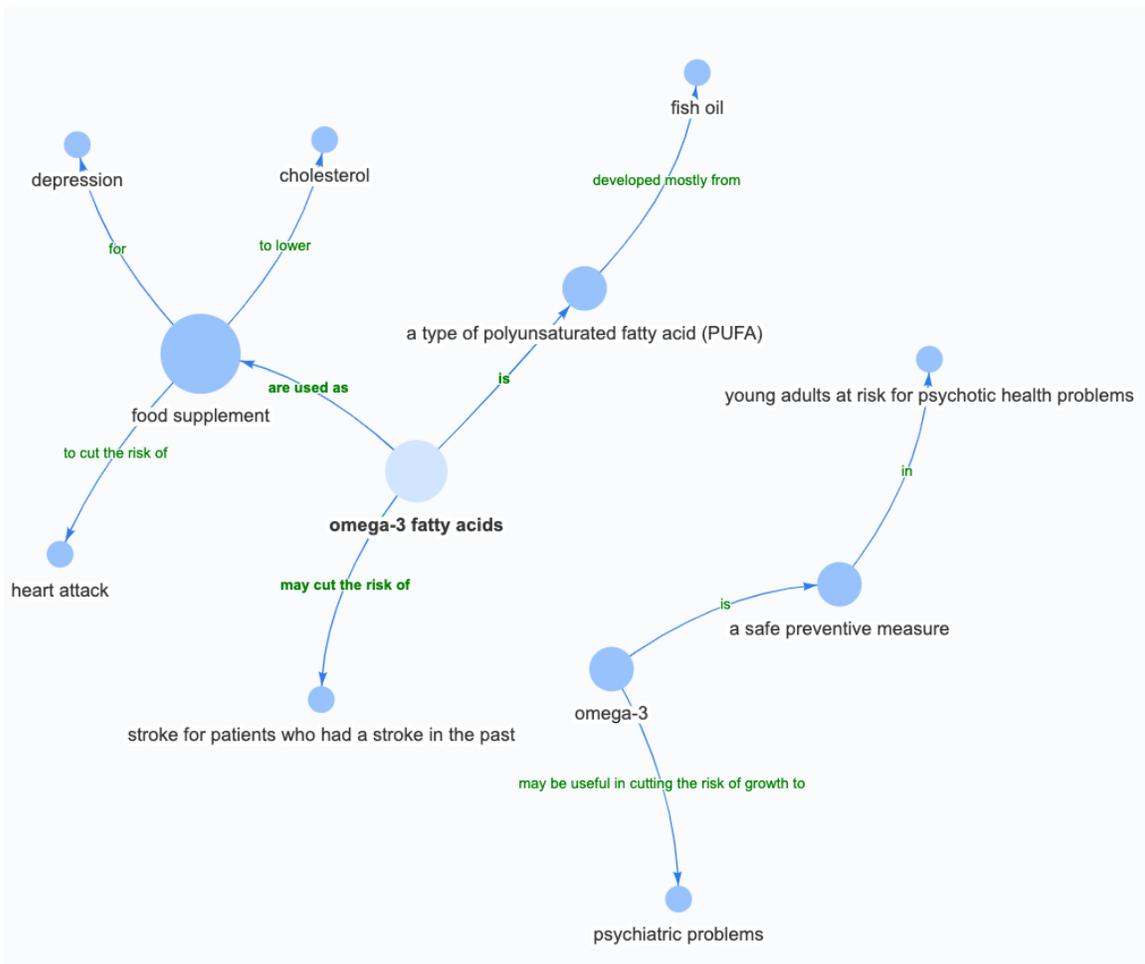

**Figure 2.** An example of the graph-based representation for the background section of "Omega 3."

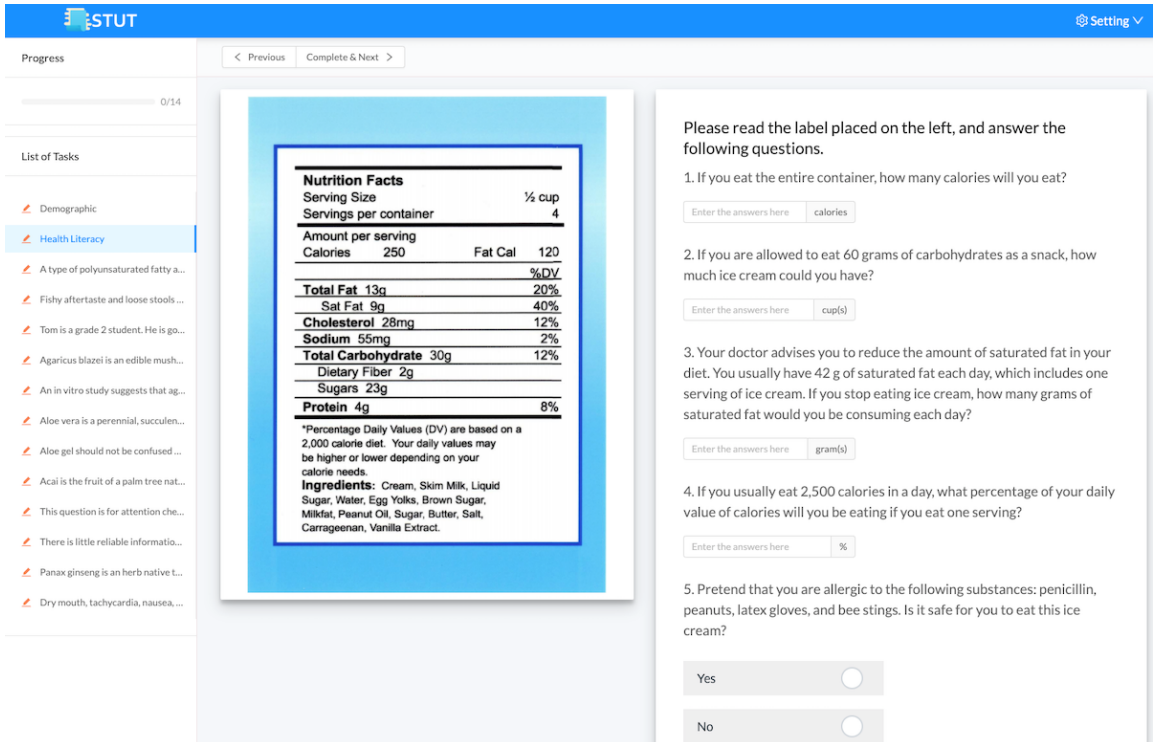

**Figure 3.** Health literacy assessment using the Newest Vital Sign assessment tool.

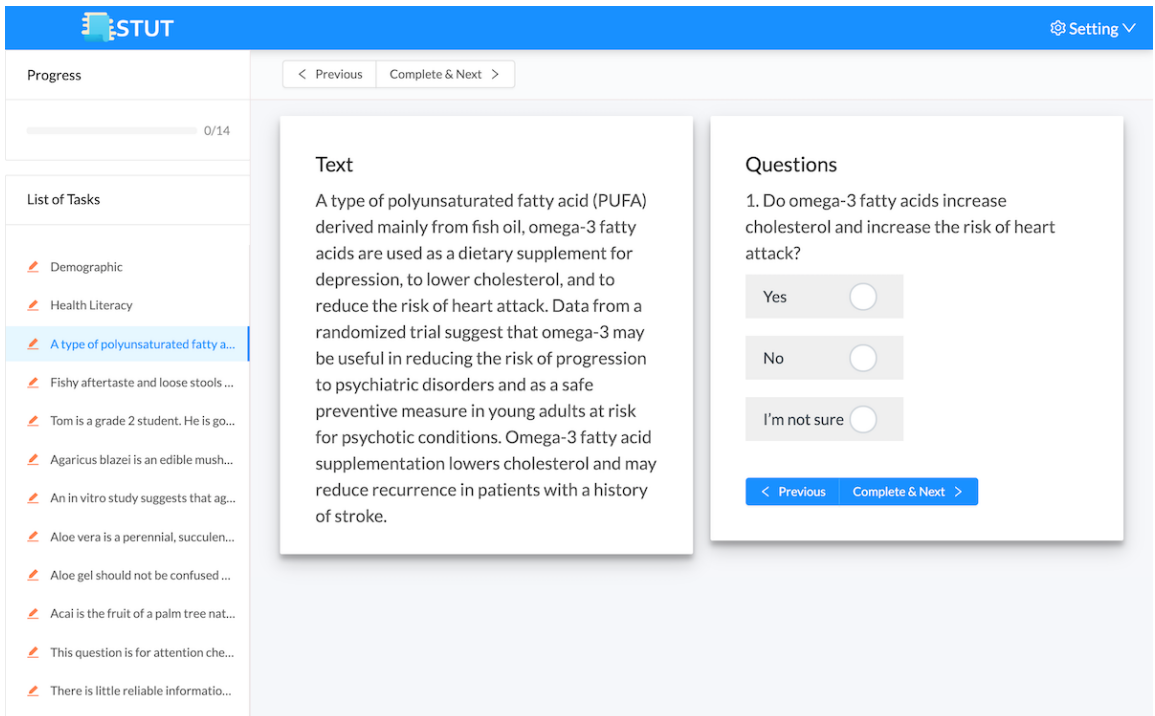

**Figure 4.** A screenshot of the user interface of the Simplified Text Understanding Test (STUT) tool.

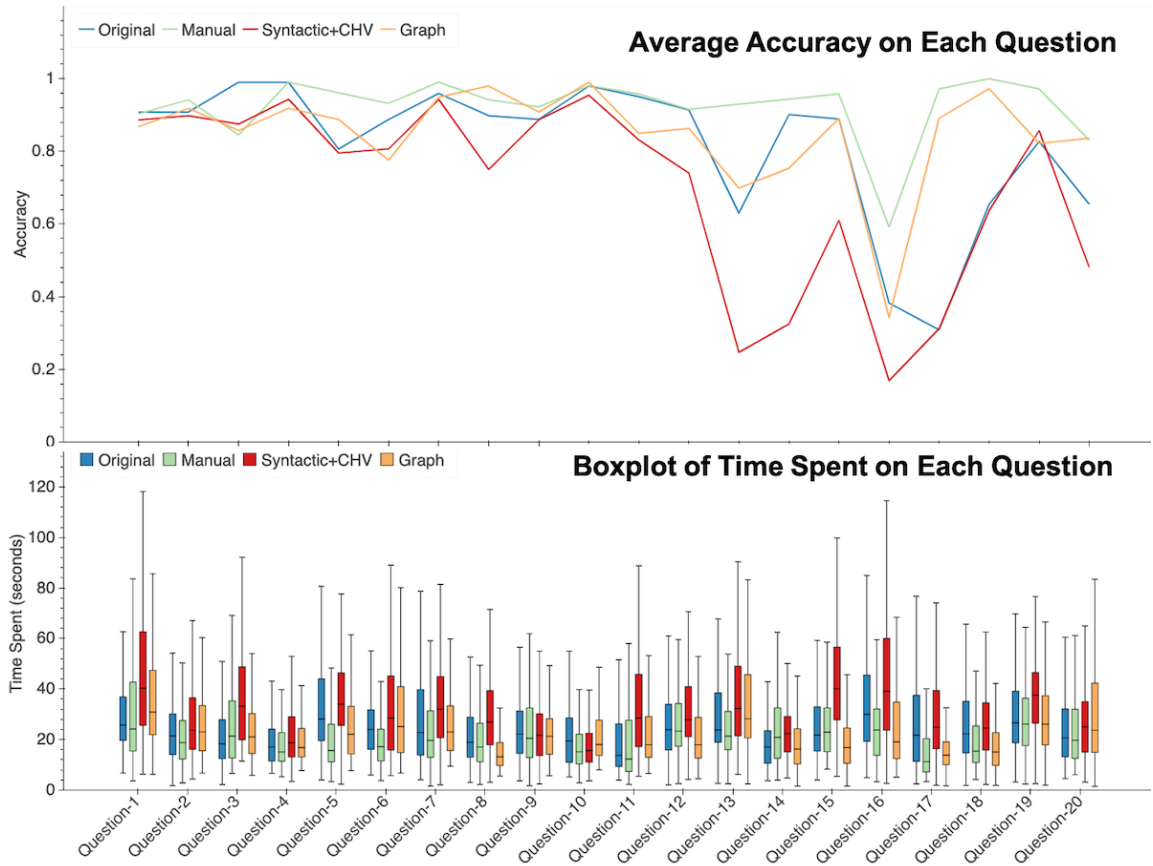

**Figure 5.** Average rates of correct answers and average time spent on each individual question across the four representations.